\documentclass[journal]{IEEEtran}
\usepackage{graphicx}
\usepackage{amsmath}

\begin{document}
\title{Re-engineering a nanodosimetry Monte Carlo code into Geant4: software design and first results}
%
%

\author{E. Gargioni, V. Grichine and M. G. Pia
\thanks{Manuscript received November 18, 2009.} 
\thanks{Elisabetta Gargioni is with University Medical Center 
	Hamburg-Eppendorf, Germany.}
\thanks{Vladimir Grichine is with Lebedev Institute, Moscow, Russia and
	CERN, Geneva, Switzerland.}
\thanks{Maria Grazia Pia is with INFN Sezione di Genova, Genova, Italy.}%
}

\maketitle
\pagestyle{empty}
\thispagestyle{empty}

\begin{abstract}
A set of physics models for nanodosimetry simulation is being
re-engineered for use in Geant4-based simulations.
This
extension of Geant4 capabilities is part of a larger scale R\&D project
for multi-scale simulation involving adaptable, co-working condensed
and discrete transport schemes. The project in progress reengineers the
physics modeling capabilities associated with an existing FORTRAN  
track-structure code for nanodosimetry into a software design suitable 
to collaborate with an object oriented simulation kernel. 
The first experience and results of the ongoing re-engineering process 
are presented.
\end{abstract}


\section{Introduction}
\IEEEPARstart{M}{ethods} 
to model hard interactions of particles with matter constituents by
means of an appropriate binary theory are well established: in this
approach collisions are treated as binary processes, that is, either
the target electrons are treated as free and at rest, or the influence
of binding is accounted for only in an approximated way.

General-purpose Monte Carlo codes, like EGS \cite{egs4,egsnrc,egs5}, 
FLUKA \cite{fluka1,fluka2},
Geant4 and MCNP \cite{mcnp,mcnp5,mcnpx},
operate in this context. Their calculations of energy deposit
distributions are based on condensed-random-walk schemes of particle
transport; this approach is adequate as long as the discrete energy
loss events treated are of magnitudes larger than electronic binding
energies.

Various specialized Monte Carlo codes, usually known as "track
structure codes", have been developed for microdosimetry and nanodosimetry
calculations. They handle particle interactions with matter as
discrete processes: all collisions are explicitly simulated as
single-scattering interactions. This approach is suitable to studies
where the precise structure of the energy deposit and of the secondary
particle production associated with a track is essential;
nevertheless, the detailed treatment of collisions down to very low
energy results in a high computational demand.

There is increasing evidence that the pattern of radiation
interaction on the nanometer level is critical for the biological
effects of ionizing radiation; in addition, radiation effects at the
nano-scale are important for the protection of electronic devices
operating in various radiation environments. 

In realistic use cases
such small-scale systems are often embedded in larger scale ones: for
instance, a component may operate within a HEP experiment or on a
satellite in space, cellular and sub-cellular aggregates in real
biological systems exist in complex body structures etc. 

So far, simulation based on condensed-random-walk schemes and track
structure generation have been treated as distinct computational
domains. 
The separation of the two simulation domains is due to the conceptual and
technical difficulty of handling the two schemes in the same
simulation environment. Achieving a conceptual approach and an
architectural design where the two schemes can co-work would represent
a significant progress in Monte Carlo simulation.

Recently, a set of specialized processes for track structure
simulation in liquid water has been designed and implemented in
Geant4 \cite{dna}; like their equivalent in dedicated Monte Carlo codes, they
operate in the r\'egime of discrete interactions only.
As a further step to provide the experimental community with tools 
for Geant4-based simulations at the nano-scale, the physics capabilities 
of a Monte Carlo code for nanodosimetry simulation
\cite{grosswendt}  are being re-engineered for use with Geant4 kernel.
This extension of Geant4 capabilities is part of a larger scale R\&D
project for multi-scale simulation, which investigates the concept of
adaptable, co-working condensed and discrete transport schemes 
\cite{mc2009,nano5}.

\section{Nanodosimetric distributions}

Nanodosimetric quantities, like size distributions of clustered
ionization, are important in understanding the radiation-induced
damage in biological targets of nanometric size (such as DNA segments)
\cite{grosswendt2}. 
These quantities, however, are not directly measurable in
biological targets and their actual knowledge is mostly based on
theoretical models. A practice to overcome this problem is to measure
cluster-size distributions using a nanodosimeter, which consists
basically of a gas-filled counter operating at low pressure. 

A typical nanodosimeter, as developed in several institutions in the
past few years \cite{garty,denardo,bantsar}, 
consists of a low-pressure interaction chamber, an
electrode system to extract ions or low-energy electrons from the
interaction chamber, an evacuated drift column which includes at its
end a single-particle counter, and a primary-particle detector. When
charged particles enter the interaction chamber, they penetrate
through or pass aside a wall-less target volume of definite shape and
size, and finally reach a trigger detector at the opposite end of the
chamber. Ions or low energy electrons produced within the target
volume are extracted from the interaction chamber and guided into an
evacuated drift chamber where they are detected by a single-particle
counter. If such measurements are performed for a large number of
primary particles of radiation quality Q, the final result is the
probability distribution $P_{n}(Q)$ of ionization cluster size \textit{n}.

These nanodosimeters were developed aided by
the important contribution of particle-track Monte Carlo
simulations. These simulations require an accurate, complete and
consistent set of scattering cross sections in the gas of interest. 

In order to take advantage of the geometrical modeling capabilities and
the features for the description of larger scale environments
available in multi-purpose Monte Carlo codes, it is desirable to
integrate their capabilities with an appropriate particle-track code,
capable of accurate transport of electrons with kinetic energies below
1 keV, down to the ionization threshold.

\section{Reengineering process}

The project in progress aims at reengineering the physics modeling 
capabilities of a track-structure code developed in the past years
for nanodosimetry applications \cite{grosswendt} into a set of simulation
tools compatible with Geant4 kernel.
The corresponding software process is based on the Unified Process \cite{up}: 
it adopts an iterative and incremental life-cycle, providing concrete 
deliverables at each iteration.

The original ``track structure'' code was developed according to a
procedural programming paradigm and is written in FORTRAN.
The availability of its functionality for operation in an open source,
general purpose simulation environment would provide experimentalists
more powerful modeling capabilities for nanodosimetric studies than
the original code: researchers would be able to exploit the rich
functionality of other Geant4 domains, like geometry, navigation,
visualization and other features, along with the specialized physics
available in the original standalone FORTRAN code.

Once reengineered for use compatible with Geant4 kernel, the functionality
corresponding to ``track structure'' simulation, i.e. pertinent to a discrete
transport scheme, would provide a valuable playground for the ongoing
research on co-working condensed and discrete schemes, which is the object
of the Nano5 \cite{nano5} R\&D process.

The introduction of equivalent physics capabilities in an object oriented
simulation environment is directly based on the literature
documenting the physics models implemented in the original track structure
system, rather than on the FORTRAN code itself.
The largely different software technology adopted by the two systems
prevents the reuse of the existing procedural code: a sound solution
requires 
rethinking the software design supporting the physics capabilities 
in terms of an object oriented paradigm.

Therefore the whole reengineering process is performed on the ground
of the pertinent physics literature: the problem domain analysis, the
software design and the implementation.
This process ensures the optimization of the software quality
through the adoption of established best practices, and the effective 
investment of the limited available resources.

The reengineering process is currently going through the inception phase.
Requirements are captured from the pertinent literature and analyzed; their 
impact on Geant4 simulation kernel is evaluated in depth.
At this stage the analysis and design process explores the capability of 
policy-based design \cite{alexandrescu}
of supporting the physics functionality of a large scale 
track structure code in an object oriented environment. 

The agility of the design, discussed in \cite{em}, facilitates testing
physics functionality thoroughly along with the implementation process:
the test process can easily encompass both the verification of the 
correctness of the implementation and the validation against experimental 
data (if available) at the level of unit testing.

\section{First results: electron impact ionization}

Ionization produced by slow (kinetic energy smaller than few keV)
electrons is one of the main processes which should be simulated in
nanodosimetry. 
The mean free path of electrons due to the ionization
process is defined by the total ionization cross section. 
Here we show
some preliminary results concerning the implementation of the Binary
Encounter Bethe (BEB) model for the total ionization cross section of
slow electrons with atoms and molecules.
The model has been implemented in a policy-based class design similar
to the approach adopted in \cite{em} to model photon interactions.
 
\subsection{Binary encounter Bethe model}

According to the binary encounter Bethe model \cite{kr94}, the total 
ionization cross section of 
slow electrons for an atomic shell, $i$,  with binding energy $B_i$ reads:
%

\begin{eqnarray*}
\label{beb}
\lefteqn{\sigma_i(t)=} \\
& & \frac{S_i}{t+u+1} \times \\
& &\left\{\frac{Q_i}{2}\left(1-\frac{1}{t^2}\right)\ln t+(2-Q_i)\left[\left(1-\frac{1}{t}\right)-\frac{\ln t}{t+1}\right]\right\},
\end{eqnarray*}
\[
t=\frac{T}{B_i}, \quad u=\frac{U_i}{B_i},
\]
where $S_i=4\pi a_o^2N_i(R/B_i)^2$, $R$ is the Rydberg energy
($13.61$ eV), $a_o$ is the Bohr radius ($5.29\times 10^{-11}$ cm)
and $N_i$ is the number of bound electrons in the $i$-shell. 
$T$ is the kinetic energy of the incident electron, and $U_i$ is the average
kinetic energy of the $i$-shell electrons.
The $Q_i$ value relates to the atomic oscillator strength distribution, 
$df/dw$:
\[
Q_i=\frac{2}{N_i}\int_{0}^{\infty}\frac{1}{w+1}\frac{df}{dw}dw, \quad  w=\frac{W}{B_i},
\]
where $W$ is the kinetic energy of the ejected (ionization) electron. 

In the cases, when the distribution $df/dw$ is unknown, one can set 
$Q_i=1$~\cite{ptb08}. 
Then equation (\ref{beb}) can be simplified:
\begin{equation}
\label{sbeb}
\sigma_i(t)=\frac{S_i}{t+u+1}\left\{\frac{\ln t}{2}\left(1-\frac{1}{t^2}\right)+
1-\frac{1}{t}-\frac{\ln t}{t+1}\right\}.
\end{equation}
The total  ionization cross section is defined by the 
summation of equation (\ref{sbeb}) over all involved atomic shells. 
The number of the shells, as well as the shell parameters, can be either 
parametrized for molecules \cite{hkr95}, or derived from atomic 
databases~\cite{kr94}.

\subsection{Comparison of simulation with experimental data}

\begin{figure}
\includegraphics[width=8.5cm]{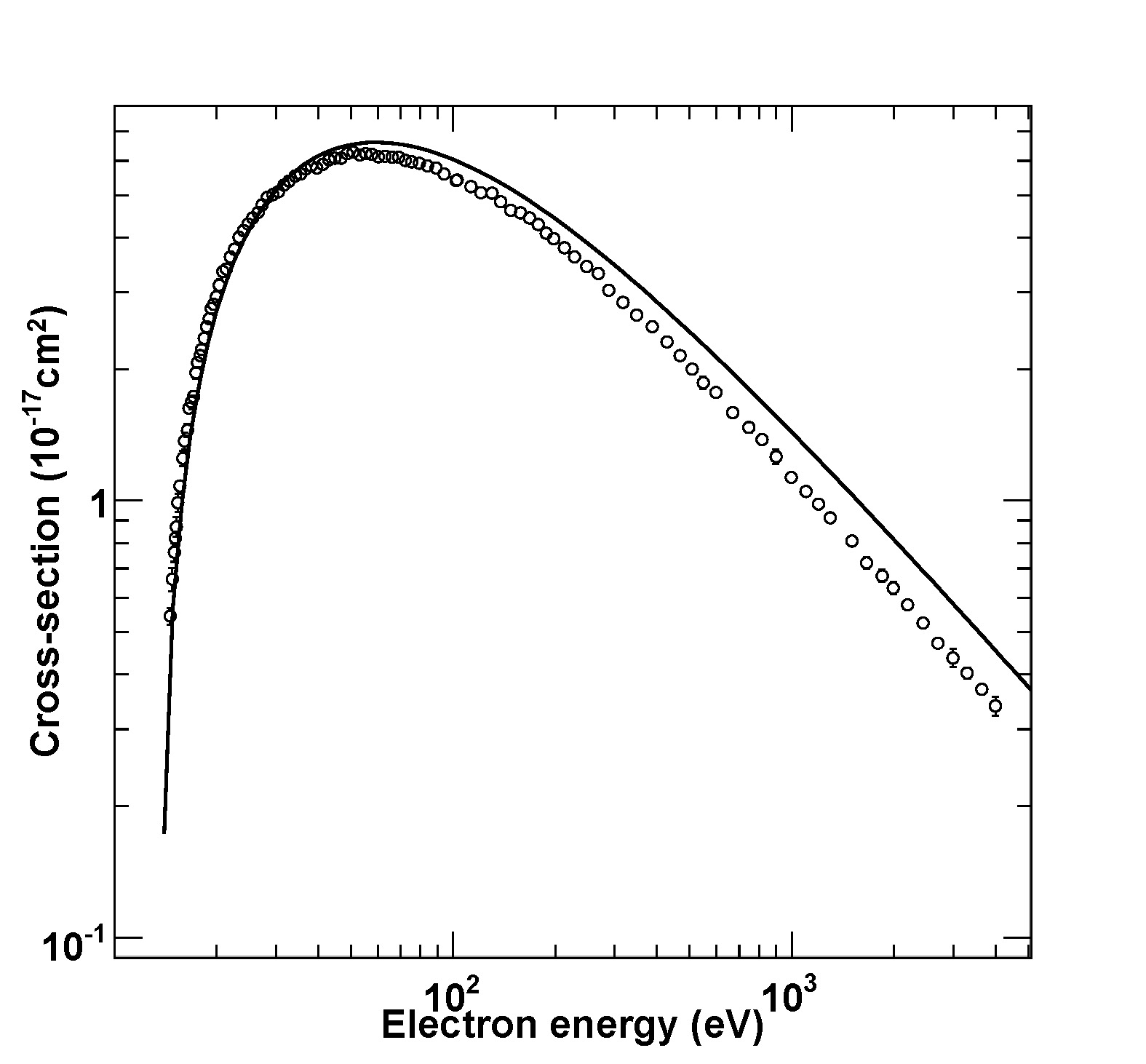}
\caption{Total cross section for ionization of atomic hydrogen by 
low energy electrons.  
Experimental data (open circles) are from \cite{shah87}, the solid line is 
our simulation according to equation (\ref{sbeb}).}
\label{eH}
\end{figure}

The total ionization cross section simulated according to the
summation of equation (\ref{sbeb}) over all involved atomic shells
was compared with experimental data for hydrogen and propane.
Fig. \ref{eH} shows the energy dependence of the ionization cross section 
of electrons on atomic hydrogen. 
Experimental data (open circles) are from \cite{shah87}, the solid line is 
our simulation according to equation (\ref{sbeb}). 

\begin{figure}
\includegraphics[width=8.5cm]{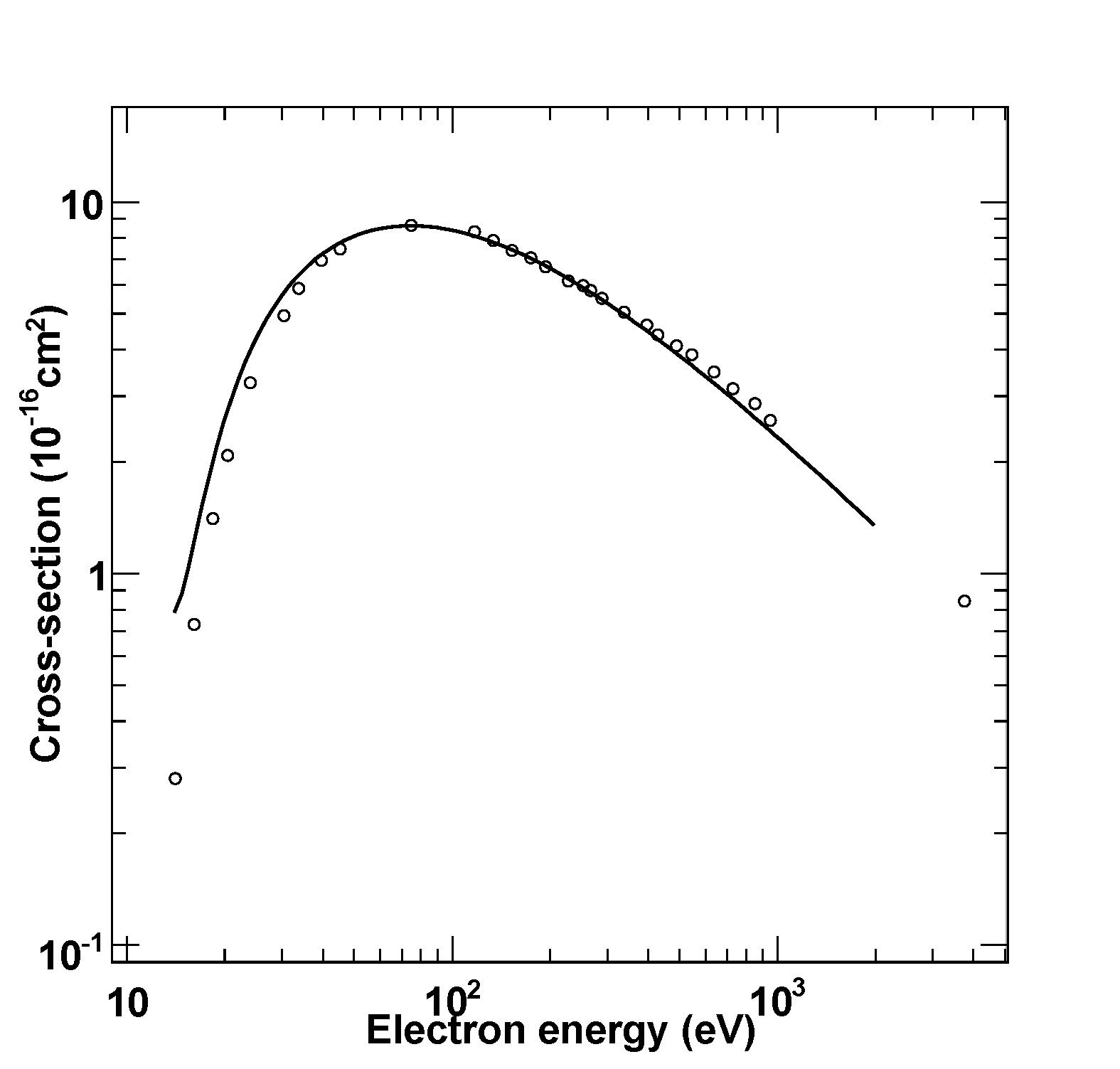}
\caption{Total cross section for ionization of $C_3H_8$ by low energy 
electrons. 
Experimental data (open circles) are from \cite{grill93}, the solid line 
is our simulation according to equation (\ref{sbeb}). 
The propane molecular parameters used in the simulation were parameterized 
in~\cite{hkr95}.
}
\label{eC3H8}
\end{figure}

Fig. \ref{eC3H8} shows the energy dependence of the electron impact 
ionization cross section for propane. 
Experimental data (open circles) are from \cite{grill93}, the solid line 
is our simulation according to equation (\ref{sbeb}). 
The propane molecular parameters used in the simulation were parameterized 
in~\cite{hkr95}.

These preliminary comparisons show satisfactory agreement of our simulation 
with experimental data. 
 
A significant metric in the reengineering 
process is the time investment for the achievement
of these results: it amounted to a few days' work, including the study
of the pertinent literature, the implementation of the physics model, the 
verification of the software and the validation against experimental data.

\section{Conclusion and outlook}
A process is in progress to reengineer the physics capabilities of an  
existing software system for nanodosimetry simulation in a software environment
compatible with Geant4 kernel; as a further step, the seamless transition of
transport schemes between nanodosimetric and conventional dosimetry simulation
will be object of investigation.

The first results of the reengineering process are encouraging: 
they document the feasibility of implementing the cross section model 
in a policy class, the achievement of satisfactory physics accuracy and
a preliminary estimate of the expected development effort.

\section*{Acknowledgment}
The authors thank 
Simone Giani (CERN), 
Bernd Grosswendt (formerly PTB, retired),
Reinhard Schulte (Loma Linda University)
and Andrew Wroe (Loma Linda University)
for helpful discussions.

\end{document}